\documentclass{aa}

\usepackage{graphicx}
\usepackage{txfonts}
\usepackage[usenames,dvipsnames]{xcolor}

\begin{document} 

\title{Molecular Line Emission as a Tool for Galaxy Observations (LEGO)}

\subtitle{I. HCN as a tracer of moderate gas densities in molecular clouds and galaxies}

\author{Jens Kauffmann\inst{1}
  \and
  Paul F.\ Goldsmith\inst{2}
  \and
  Gary Melnick\inst{3}
  \and
  Volker Tolls\inst{3}
  \and
  Andres Guzman\inst{4}
  \and
  Karl M.\ Menten\inst{1}
}

\institute{Max--Planck--Institut f\"ur Radioastronomie, Auf dem H\"ugel 69, D--53121 Bonn, Germany, \email{jens.kauffmann@gmail.com}
  \and
  Jet Propulsion Lab, California Institute of Technology, 4800 Oak Grove Drive, Pasadena, CA 91109, USA
  \and
  Harvard--Smithsonian Center for Astrophysics, 60 Garden Street, Cambridge, MA 02138, USA
  \and
  Departamento de Astronom\'ia, Universidad de Chile, Camino el Observatorio 1515, Las Condes, Santiago, Chile
}

\date{received XXX; accepted XXX}

 
\abstract{Trends observed in galaxies, such as the Gao \& Solomon relation, suggest a linear relation between the star formation rate and the mass of dense gas available for star formation. Validation of such relations requires the establishment of reliable methods to trace the dense gas in galaxies. One frequent assumption is that the HCN ($J=1$--0) transition is unambiguously associated with gas at $\rm{}H_2$ densities $\gg{}10^4~\rm{}cm^{-3}$. If so, the mass of gas at densities $\gg{}10^4~\rm{}cm^{-3}$ could be inferred from the luminosity of this emission line, $L_{\rm{}HCN\,(1\text{--}0)}$. Here we use observations of the Orion~A molecular cloud to show that the HCN ($J=1$--0) line traces much lower densities $\sim{}10^3~\rm{}cm^{-3}$ in cold sections of this molecular cloud, corresponding to visual extinctions $A_V\approx{}6~\rm{}mag$. We also find that cold and dense gas in a cloud like Orion produces too little HCN emission to explain $L_{\rm{}HCN\,(1\text{--}0)}$ in star--forming galaxies, suggesting that galaxies might contain a hitherto unknown source of HCN emission. In our sample of molecules observed at frequencies near 100~GHz (also including $\rm{}^{12}CO$, $\rm{}^{13}CO$, $\rm{}C^{18}O$, CN, and CCH), $\rm{}N_2H^+$ is the only species clearly associated with rather dense gas.}


\maketitle


\section{Introduction\label{sec:introduction}}
The relationship between star formation (SF) and the supply of dense gas is of critical importance for our understanding of cosmic SF. We must develop a detailed picture of the relation between dense gas and SF in galaxies if we wish to explain the structure and evolution of galaxies (e.g., \citealt{somerville2014:sf-cosmology-simulations}). This relation can, for example, be explored in the Milky Way. In molecular clouds within $\sim{}500~\rm{}pc$ from Sun one can estimate the star formation rate, $\dot{M}_{\star}$, by counting individual young stars. These nearby clouds can be resolved spatially, which also simplifies estimating the mass of gas at high density, $M_{\rm{}dg}$. Recent research suggests defining $M_{\rm{}dg}$ as the mass residing at high visual extinctions, $A_V\ge{}A_{V,{\rm{}dg}}$ with $A_{V,\rm{}dg}\approx{}7~\rm{}mag$, resulting in $\dot{M}_{\star}\propto{}M_{\rm{}dg}$ (e.g., \citealt{heiderman2010:sf-law}, \citealt{lada2010:sf-efficiency}).

It is very challenging to study $\dot{M}_{\star}$ and $M_{\rm{}dg}$ in galaxies. One might, for example, assume that the light of young stars is absorbed and re--emitted by dust. Then the far--infrared luminosity of a galaxy (i.e., at wavelengths of 8 to $1,000~\rm{}\mu{}m$) characterizes SF via $\dot{M}_{\star}\propto{}L_{\rm{}FIR}$. Similarly, one might assume that a certain molecular emission line requires elevated densities to be excited. Then $M_{\rm{}dg}\propto{}L_Q$ for line luminosities of a suitable transition $Q$. \citet{gao2004:hcn}, in particular, introduced the HCN ($J=1$--0) transition as a tracer of dense gas in galaxies (i.e., $\rm{}H_2$ densities $\gg{}10^4~\rm{}cm^{-3}$), suggesting that $M_{\rm{}dg}\propto{}L_{\rm{}HCN\,(1\text{--}0)}$.

This raises an important question: is $M_{\rm{}dg}$ as derived from $A_V$ equal to $M_{\rm{}dg}$ as obtained from $L_{\rm{}HCN\,(1\text{--}0)}$? The LEGO project (Molecular Line Emission as a Tool for Galaxy Observations; led by JK) uses wide--field maps to address such questions. We here summarize key conclusions from a comprehensive study of Orion~A (Kauffmann et al., in prep.; hereafter Paper~II).

\begin{figure*}
\centerline{\includegraphics[width=\linewidth]{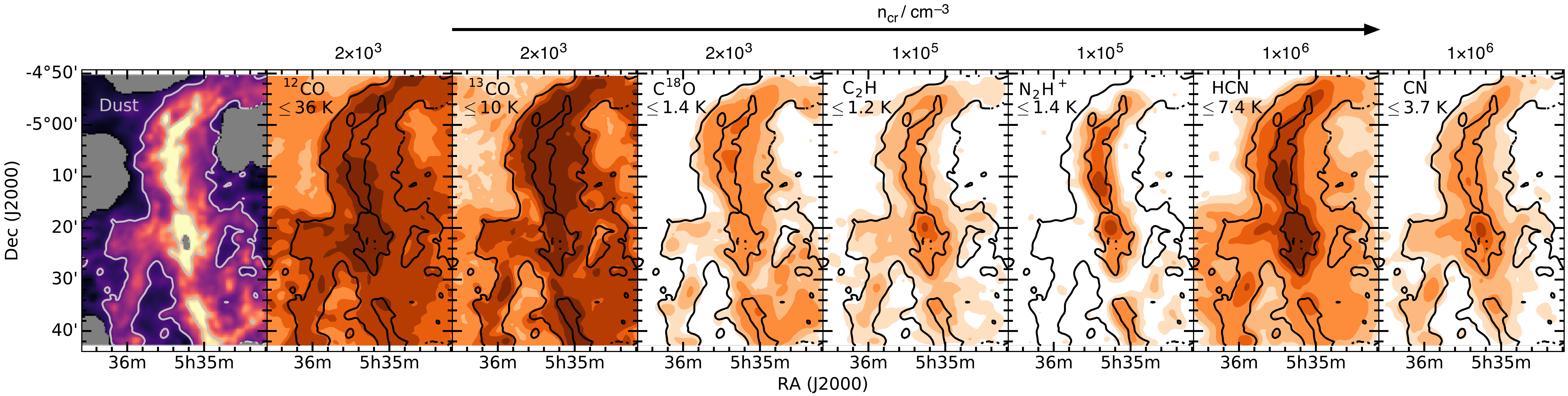}}
\caption{Maps of the peak intensity for transitions near 100~GHz. The left panel gives $A_V$ as inferred from Herschel data. Contours at 5 and 30~mag are drawn and repeated in all panels, and the peak intensity is stated for every transition. Line emission maps are smoothed to $1\farcm{}5$ resolution before filled contours are drawn at signal--to--noise ratios of 3, 5, 10, 30, 50, and 100. Panels are ordered by increasing critical density, $n_{\rm{}cr}$.\label{fig:int_peak}}
\end{figure*}

\section{Preparation of Observational Data}
Data on emission lines at $\sim{}100~\rm{}GHz$ frequency (Fig.~\ref{fig:int_peak}) were obtained with the 14m--telescope of the Five College Radio Astronomy Observatory (FCRAO). Maps of the CCH ($N=1$--0, $J=1/2$--1/2), HCN ($J=1$--0), $\rm{}N_2H^+$ ($J=1$--0), $\rm{}C^{18}O$ ($J=1$--0), and CN ($N=1$--0, $J=3/2$--1/2) transitions are taken from \citet{melnick2011:molecules-orion}. Data on the $\rm{}^{12}CO$ ($J=1$--0) and $\rm^{13}CO$ ($J=1$--0) lines are from \citet{ripple2013:fcrao-orion}. The full--width at half--maximum beam size for given frequency $\nu$ is $\vartheta_{\rm{}beam}=52\arcsec{}\cdot{}(\nu/{100~\rm{}GHz})^{-1}$. An efficiency $\eta_{\rm{}mb}=0.47$ is used for conversion  to the main beam intensity scale, $T_{\rm{}mb}=T_{\rm{}A}^{\ast}/\eta_{\rm{}mb}$. This paper focuses on the integrated intensities, $W=\int{}T_{\rm{}mb}\,{\rm{}d}v$.

Dust--based estimates of the $\rm{}H_2$ column density $N({\rm{}H_2})$ are derived from Herschel observations of Orion at wavelengths of 250 to $500~\rm{}\mu{}m$ \citep{andre2010:herschel-gb} using modified methods from \citet{guzman2015:herschel-malt90} described in \citet{kauffmann2016:gcms_i}. We assume thin ice coatings and dust coagulation for $10^5~\rm{}yr$ at a molecular volume density of $10^6~\rm{}cm^{-3}$ to select dust opacities from \citet{ossenkopf1994:opacities}. Paper~II describes how we calibrate these data against an extinction--based map from \citet{kainulainen2011:confinement} to predict the visual extinction, $A_V/{\rm{}mag}=N({\rm{}H_2})/9.4\times{}20^{20}~{\rm{}cm^{-2}}$, at a resolution of $38\arcsec$.

We fit the filamentary cloud north of $-$5:14:00 (J2000) with a
truncated cylindrical power--law density profile,
$n(r)=n_R\cdot{}(r/R)^{-k}$, where $r$ is the distance from the
filament's main axis. We then obtain the median density along any line
of sight for an offset $s$ from the filament main axis,
$n_{\rm{}med}(s)$. For given $s$, half of the mass resides above (and half below) this density, so that $n_{\rm{}med}(s)$ can be considered a representative density. Further algebraic operations relate $s$, $n_{\rm{}med}(s)$, and $A_V(s)$ (Fig.~\ref{fig:cumulative-fraction-emission}; see Paper~II).

\section{Molecules as Tracers of Cloud Material\label{sec:lines-as-tracers}}
We seek to explore molecular line emission under conditions that are representative for the Milky Way. We therefore ignore the region south of $-$5:10:00 declination (J2000). First, much of this region is subject to intense radiation emitted by young stars in the Orion Nebula. This is probably not typical for molecular clouds. Second, the well--shielded southerly regions (with dust temperatures $\le{}22~\rm{}K$) are devoid of embedded stars that are characteristic of SF regions \citep{megeath2012:orion}. Finally, we ignore pixels where $A_V<2~\rm{}mag$ because of observational uncertainties.

\begin{figure}
\includegraphics[width=\linewidth]{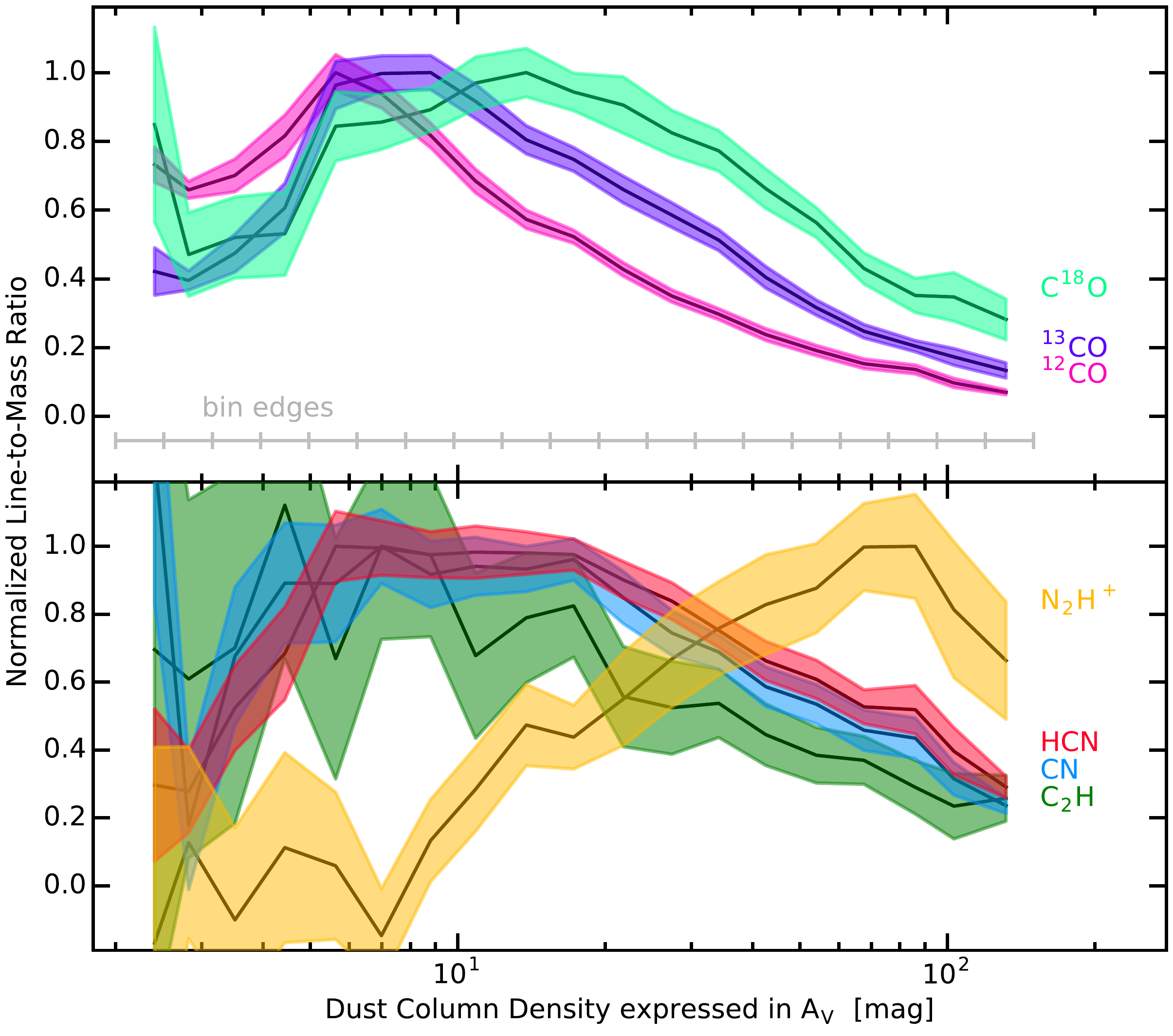}
\caption{Normalized line--to--mass ratio, $h_Q$, for the reference region. Shading indicates the uncertainty at a confidence level $\approx{}68\%$, while gray dashes indicate the limits of bins. $\rm{}N_2H^+$ is a good tracer of dense gas since $h_Q$ increases with increasing $A_V$.\label{fig:line-to-mass}}
\end{figure}

\subsection{Line Emission per Unit Cloud Mass\label{sec:line-per-mass}}
The line--to--mass ratio, $h_Q=W(Q)/A_V$, indicates how the emission from transition $Q$ relates to the mass reservoir characterized by $A_V$. Given $A_V\propto{}N({\rm{}H_2})$, the ratio $W(Q)/A_V$ essentially measures the intensity of line emission per $\rm{}H_2$ molecule.

It is plausible to assume that $h_Q$ is a function of $n$ and therefore $N({\rm{}H_2})$. This is, for example, expected if the molecular abundance or the excitation is a function of the density. This is explored in Fig.~\ref{fig:line-to-mass}. For this analysis we sort the data into logarithmically spaced bins in $A_V$, and we then derive the mean of $h_Q$ and its uncertainty from counting statistics in this bin. We see that $h_Q$ is indeed a strong function of $A_V$, which justifies our ansatz to explore the trend of $h_Q$ versus $A_V$. We normalize $h_Q$ to a maximum value of 1 in the well--detected bins of Fig.~\ref{fig:line-to-mass} in order to simplify comparisons between molecular species.

The trend of $h_Q$ vs.\ $A_V$ is non--trivial, and it differs between molecules. Most molecules start with a significant value of $h_Q$ at low $A_V$, their line--to--mass ratio increases towards a maximum at an $A_V$ of 5 to 20~mag, and $h_Q$ steadily decreases with increasing $A_V$ at even higher extinction. One single molecule defies this trend: the line--to--mass ratio of $\rm{}N_2H^+$ begins near or at zero at low $A_V$, and $h_Q$ then begins to steadily rise at $A_V\gtrsim{}10~\rm{}mag$, possibly to level out (or decrease) at $A_V\gtrsim{}100~\rm{}mag$.

This is a critical result. This means that the $\rm{}N_2H^+$ ($J=1$--0) transition is the only transition among those observed here that selectively traces gas at high (column) density. All other transitions are, by contrast, most sensitive to material at $A_V\sim{}10~\rm{}mag$. \citet{pety2016:orion-b} conclude the same in Orion~B, using an argument that relates more to our next section.

\subsection{Characteristic (Column) Density traced by a Line\label{sec:cd-characteristic}}
\begin{figure}
\includegraphics[width=\linewidth]{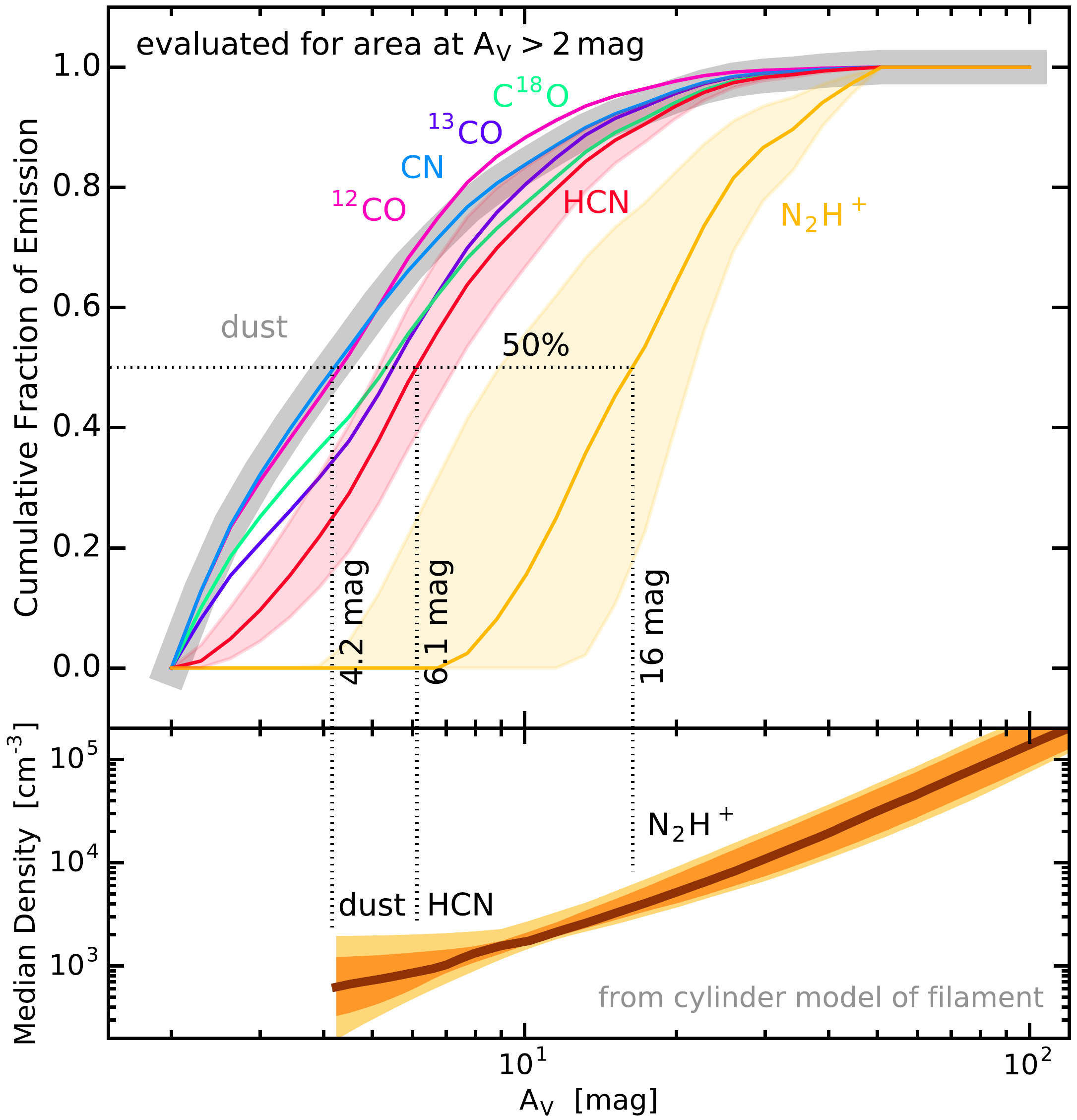}
\caption{The top panel shows the cumulative fraction of emission for various transitions (and mass for dust) indicated by various colors. Dashed vertical lines indicates where selected transitions achieve $L_Q^{\ast}=L_Q/2$. Shaded regions indicate uncertainties as described in Sec.~\ref{sec:cd-characteristic}. The bottom panel indicates the estimated median density. Shading indicates confidence ranges of $\pm{}10\%$ and $\pm{}40\%$ around the median estimate.\label{fig:cumulative-fraction-emission}}
\end{figure}

Figure~\ref{fig:line-to-mass} characterizes whether a given molecular emission line traces the cloud material well under given conditions. It would be desirable if this information could be collapsed into a single number. One could, for example, attempt to establish the typical $\rm{}H_2$ (column) density of material that is traced by a given transition. We use the line luminosities for this purpose. Integration over the map area at column
densities corresponding to $A_V\le{}A_V^{\ast}$ gives the luminosity
as a function of the cutoff value $A_V^{\ast}$,
\begin{equation}
L_Q^{\ast}(A_V^{\ast})=\int_{A_V<A_V^{\ast}}W_Q\,{\rm{}d}\mathcal{A}\,,
\label{eq:line-luminosity}
\end{equation}
where ${\rm{}d}\mathcal{A}$ is the area element measured in $\rm{}pc^2$. Let $L_Q=L_Q^{\ast}(A_V^{\ast}\to{}\infty)$ be the total luminosity. We then define the characteristic column density $A_{V,{\rm{}char}}^Q$ of transition $Q$ to be the column density that contains half of the total line luminosity,
\begin{equation}
L_Q^{\ast}(A_{V,{\rm{}char}}^Q)=L_Q/2\,.
\label{eq:characteristic-cd}
\end{equation}
We then use the density model to define a characteristic density $n_{\rm{}char}^Q=n_{\rm{}med}(A_{V,{\rm{}char}}^Q)$. We also obtain the characteristic column density for the spatial gas mass distribution, $A_{V,\rm{}char}^M$, by replacing $W_Q$ with $A_V$ in Eqs.~(\ref{eq:line-luminosity}--\ref{eq:characteristic-cd}). Figure~\ref{fig:cumulative-fraction-emission} shows how $L_Q^{\ast}(A_V^{\ast})/L_Q$ and $n_{\rm{}med}$ increase towards deeper layers of the cloud.

This analysis is greatly influenced by the choice of the region analyzed. For example, it is generally known that most of a cloud's mass resides at low column densities. The field north of $-$5:10:00 declination considered here does, however, not fulfill this condition. For example, $A_{V,\rm{}char}^M=17~\rm{}mag$ in the area shown in Fig.~\ref{fig:int_peak}, while $A_{V,\rm{}char}^M=2.0~\rm{}mag$ holds for the entire Orion~A cloud. We therefore use the $A_V$ map from \citet{kainulainen2011:confinement} to correct for this relative lack of lower--density material. Specifically, we interpolate the binned information on $h_Q$ (i.e., $W[Q]$ vs.\ $A_V$) summarized in Fig.~\ref{fig:line-to-mass} to derive a predicted value of $W(Q)$ for every pixel in the \citeauthor{kainulainen2011:confinement} map. We then use these predicted $W(Q)$ in Eqs.~(\ref{eq:line-luminosity}--\ref{eq:characteristic-cd}) to derive predictions of $L_Q^{\ast}(A_V^{\ast})/L_Q$ and $A_{V,{\rm{}char}}^Q$ for the entire Orion~A region. We do not treat CCH because of overly large uncertainties in $h_Q$.

We use the observed values of $h_Q$ from Fig.~\ref{fig:line-to-mass} if the measurements exceed their uncertainty by a factor $\ge{}3$. Around this detection limit we use linear fits to $h_Q$ vs.\ $\lg(A_V)$ to determine the $A_V$ for which $h_Q=0$. We then use linear interpolation in $h_Q$ between the point where the transition is safely detected and the point where the emission is predicted to vanish. The latter point has an uncertainty resulting from the aforementioned linear fit. Variation of this point changes $L_Q^{\ast}$ and thereby $A_{V,{\rm{}char}}^Q$. Further, one might assume that the actual mass distribution (i.e., ${\rm{}d}M/{\rm{}d}A_V$) might actually deviate from the one derived from the \citeauthor{kainulainen2011:confinement} map. Here we explore a scenario in which we vary ${\rm{}d}M/{\rm{}d}A_V$ by a factor 2 up and down at $A_V=2~\rm{}mag$, leave ${\rm{}d}M/{\rm{}d}A_V$ unchanged at $A_V\ge{}10~\rm{}mag$, and interpolate linearly at intermediate $A_V$. Figure~\ref{fig:cumulative-fraction-emission} shows how extremes in both these modifications might influence the results for HCN and $\rm{}N_2H^+$.

Figure~\ref{fig:cumulative-fraction-emission} recovers the trends already seen in Fig.~\ref{fig:line-to-mass}: most transitions trace lower--density material and have $A_{V,{\rm{}char}}^Q\approx{}(5\pm{}1)~\rm{}mag$, where $A_{V,{\rm{}char}}^Q=6.1_{-1.0}^{+1.2}~\rm{}mag$ for HCN ($J=1$--0). The only exception is $\rm{}N_2H^+$ ($J=1$--0) with $A_{V,{\rm{}char}}^Q=16_{-7}^{+5}~\rm{}mag$. This transition is the only true tracer of higher column densities. \citet{pety2016:orion-b} studied in Orion~B how cloud sectors at  different $A_V$ contribute to $L_Q$. They do not calculate $A_{V,{\rm{}char}}^Q$, but their results seem broadly consistent with ours.

\section{Tracing the Dense Gas in Star--Forming Galaxies}
\subsection{HCN as a Tracer of Moderately Dense Gas\label{sec:hcn-density}}
The constraints on $h_Q=W(Q)/A_V$ and $A_{V,{\rm{}char}}^Q$ are of critical importance for the study of star--forming galaxies. For example, \citet{gao2004:hcn} speculate that gas at densities $\gtrsim{}3\times{}10^4~\rm{}cm^{-3}$ is traced by emission in the HCN ($J=1$--0) transition. More recently, \citet{usero2015:gas-across-galaxies} assumed threshold densities as large as $3\times{}10^5~\rm{}cm^{-3}$ (they deem $10^{\text{4 to 5}}~\rm{}cm^{-3}$ likely), while \citet{jimenez-donaire2016:effective-densities} estimate threshold densities $\ge{}5\times{}10^5~\rm{}cm^{-3}$ from $\rm{}H^{13}CN$--to--$\rm{}H^{12}CN$ line ratios in galaxies. More generally, it is often argued that the high critical density of the HCN~(1--0) line, $n_{\rm{}cr}=1\times{}10^6~\rm{}cm^{-3}$, implies that this transition traces gas of very high density. However, the analysis presented here shows that $A_{V,{\rm{}char}}^{\text{HCN (1--0)}}=6.1_{-1.0}^{+1.2}~\rm{}mag$, which suggests that $n_{\rm{}char}^{\text{HCN (1--0)}}\approx{}870_{-550}^{+1240}~\rm{}cm^{-3}$. This does not fundamentally question the interpretation of trends like the \citeauthor{gao2004:hcn} relation, but it critically affects the detailed analysis of data.

The low value of $n_{\rm{}char}^{\text{HCN (1--0)}}$ is not entirely surprising. \citet{evans1999:physical_conditions} and \citeauthor{shirley2015:n_cr} (\citeyear{shirley2015:n_cr}; also see \citealt{linke1977:hcn}) point out that HCN should become detectable at ``effective'' densities $n_{\rm{}eff}\approx{}(1~{\rm{}to}~3)\times{}10^4~{\rm{}cm^{-3}}$ for gas at 10~K and regular abundances, for which $n_{\rm{}eff}\ll{}n_{\rm{}cr}$. Further, HCN can be excited by electrons at $\rm{}H_2$ densities $\ll{}n_{\rm{}cr}$ if fractional electron abundances $X({\rm{}e^-})>10^{-5}$ prevail (\citealt{goldsmith2017:electron-excitation}, following a suggestion by S.~Glover). Here we provide solid observational evidence supporting such work.

Critical densities simply do not control how line emission couples to dense gas. This is already evident from Fig.~\ref{fig:int_peak}.

The low characteristic density $\approx{}870~\rm{}cm^{-3}$ for the HCN~(1--0) line has important implications for modeling. Theoretical studies often relate $\dot{M}_{\star}$ and $M_{\rm{}dg}$ via the free--fall time at density $n_{\rm{}char}^Q$, $\tau_{\rm{}ff}\approx{}10^5~{\rm{}yr}\cdot{}(n_{\rm{}char}^Q/10^5~{\rm{}cm^{-3}})^{-1/2}$, and a star formation efficiency, $\varepsilon_{\rm{}SF}\le{}1$, via $\dot{M}_{\star}=\varepsilon_{\rm{}SF}\cdot{}M_{\rm{}dg}/\tau_{\rm{}ff}$. A landmark paper by \citet{krumholz2007:slow-sf}, for example, assumes $n_{\rm{}char}^{\text{HCN (1--0)}}\approx{}6\times{}10^4~\rm{}cm^{-3}$, infers $\varepsilon_{\rm{}SF}\approx{}0.01$, and concludes that SF is ``slow'' in regions sampled by HCN (1--0). Our measurements indicate that $n_{\rm{}char}^{\text{HCN (1--0)}}$ is a factor $\approx{}70$ smaller, $\tau_{\rm{}ff}$ a factor $\approx{}70^{1/2}\approx{}8$ larger, and SF thus by a similar factor more efficient and ``faster''. Determinations of $n_{\rm{}char}^Q$ for HCN and other molecules are thus of essential importance for SF theory.

\begin{figure}
\includegraphics[width=\linewidth]{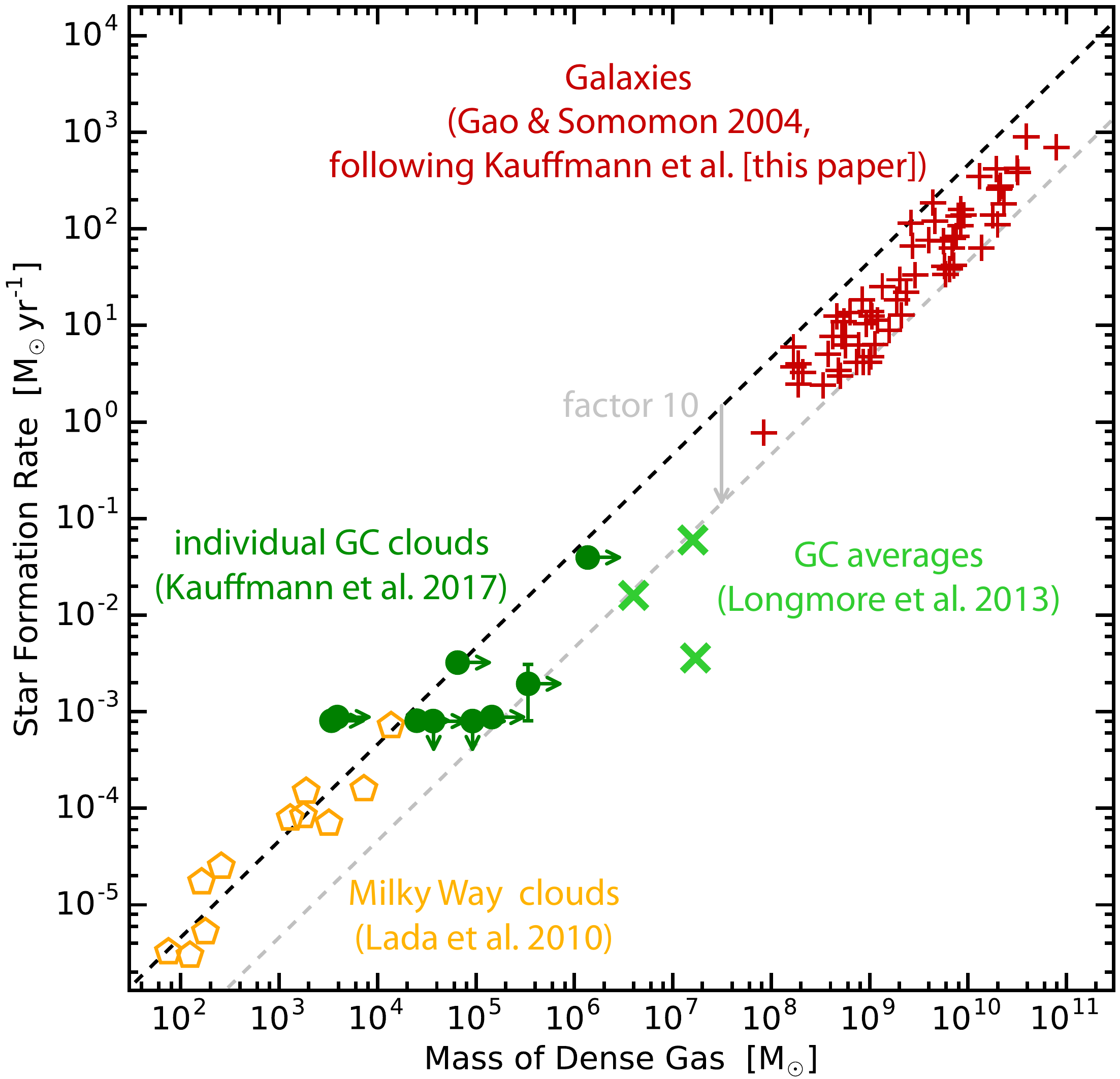}
\caption{Star formation in the Milky Way and galaxies. The reference relation for the Milky Way does not describe galaxies. This might hint at unknown reservoirs of HCN emission.\label{fig:sf-law}}
\end{figure}

\subsection{Galactic vs.\ Extragalactic Star Formation Relations\label{sec:sf-relations}}
We initially set out to investigate whether $M_{\rm{}dg}$ as derived from $A_V$ is equal to $M_{\rm{}dg}$ as obtained from $L_{\rm{}HCN\,(1\text{--}0)}$. We now return to this question. \citet{lada2010:sf-efficiency} argue that $\dot{M}_{\star}\propto{}M_{\rm{}dg}$ if $M_{\rm{}dg}$ is calculated as the cloud mass residing at $A_V\gtrsim{}A_{V,\rm{}dg}=7~\rm{}mag$. In Sec.~\ref{sec:cd-characteristic} we have shown that $A_{V,\rm{}char}^Q\approx{}6~\rm{}mag$ for the HCN (1--0) transition is similar to $A_{V,\rm{}dg}$. This suggests that about half of $L_{\text{HCN (1--0)}}$ originates directly in the dense star--forming gas of galaxies. The remaining fraction of $L_{\text{HCN (1--0)}}$ does not directly trace $M_{\rm{}dg}$. Still, this emission might be a decent probe of the gas surrounding and shaping $M_{\rm{}dg}$. From this perspective one might postulate
\begin{equation}
M_{\rm{}dg}=\alpha_{\text{HCN (1--0)}}\cdot{}L_{\text{HCN (1--0)}} \, ,
\end{equation}
where $\alpha_{\text{HCN (1--0)}}$ is a constant. \citet{gao2004:hcn}, for example, suggested that $\alpha_{\text{HCN (1--0)}}\approx{}10\,M_{\sun}/({\rm{}K\,km\,s^{-1}\,pc^2})$, based on simple models. But $\alpha_{\text{HCN (1--0)}}$ has never been estimated using observations, in particular not down to densities $<10^3~\rm{}cm^{-3}$ that we suggest are traced by the HCN (1--0) line.

We estimate $L_{\text{HCN (1--0)}}\gtrsim{}810_{-30}^{+40}~\rm{}K\,km\,s^{-1}\,pc^2$, following the procedure described in Sec.~\ref{sec:cd-characteristic}. Recall that this is a lower limit since we cannot predict $W_Q$ for $A_V<2~\rm{}mag$ (Sec.~\ref{sec:lines-as-tracers}). We further derive $M_{\rm{}dg}\approx{}1.6\times{}10^4\,M_{\sun}$ by evaluating the mass of material residing at $A_V\gtrsim{}7~\rm{}mag$ in the \citet{kainulainen2011:confinement} extinction map. We thus find $\alpha_{\text{HCN (1--0)}}\lesssim{}20\,M_{\sun}/({\rm{}K\,km\,s^{-1}\,pc^2})$. This is in good agreement with modeling by \citet{gao2004:hcn} --- but for the wrong reasons, given their models essentially assume  $n_{\rm{}char}^{\text{HCN (1--0)}}\approx{}3\times{}10^4~{\rm{}cm^{-3}}$, which exceeds the true value by a factor $\approx{}{}30$. \citet{shimajiri2017:hcn} estimate $\alpha_{\text{HCN (1--0)}}\approx{}10\,M_{\sun}/({\rm{}K\,km\,s^{-1}\,pc^2})$ from observations of Aquila, Ophiuchus, and Orion~B. Their work assumes a scaling factor to include gas at $A_V<8~\rm{}mag$. Our work differs from theirs in that we actually measure this factor (Fig.~\ref{fig:cumulative-fraction-emission}) while \citeauthor{shimajiri2017:hcn} implement a sophisticated treatment of interstellar radiation fields.

In Fig.~\ref{fig:sf-law} we use our new observational determination of $\alpha_{\text{HCN (1--0)}}$ to compare SF in the \citet{gao2004:hcn_survey} galaxies to SF in molecular clouds near the Sun \citep{lada2010:sf-efficiency} and in the Galactic Center (GC; \citealt{longmore2012:sfr-cmz}, \citealt{kauffmann2016:gcms_i}). For the galaxies we adopt $\dot{M}_{\star}=\beta_{\rm{}FIR}\cdot{}L_{\rm{}FIR}$ with $\beta_{\rm{}FIR}\approx{}3\times{}10^{-10}\,M_{\sun}\,{\rm{}yr^{-1}}\,L_{\sun}^{-1}$ (Eq.~[4] and the offset from Fig.~3 of \citealt{murphy2011:sf-calibration}). \citet{lada2010:sf-efficiency} suggest a reference relation describing SF rates in clouds within $\sim{}500~\rm{}pc$ from Sun, $\dot{M}_{\star,\rm{}MW}=(4.6\pm{}2.6)\times {}10^{-8}\,M_{\sun}\,{\rm{}yr^{-1}}\cdot{}(M_{\rm{}dg} / M_{\sun})$, from which GC clouds appear to deviate by a factor $\sim{}10$.

Figure~\ref{fig:sf-law} shows that also galaxies deviate from $\dot{M}_{\star,\rm{}MW}$ by an average factor $\langle{}\dot{M}_{\star,\rm{}MW}/\dot{M}_{\star}\rangle{}\lesssim{}4.5$. Given that $\langle{}\dot{M}_{\star,\rm{}MW}/\dot{M}_{\star}\rangle{}\propto{}\alpha_{\text{HCN (1--0)}}$, could $\alpha_{\text{HCN (1--0)}}$ in galaxies be smaller than estimated here? Significant contributions to $L_{\text{HCN (1--0)}}$ from reservoirs outside those considered here, for example from diffuse cloud envelopes, could indeed reduce $\alpha_{\text{HCN (1--0)}}=M_{\rm{}dg}/L_{\text{HCN (1--0)}}$.

\section{Summary}
We study the relationship between various emission lines and dense gas. This analysis is based on observations of various molecules at frequencies near 100~GHz ($\rm{}^{12}CO$, $\rm{}^{13}CO$, $\rm{}C^{18}O$, CN, CCH, HCN, and $\rm{}N_2H^+$). We focus on the HCN (1--0) transition, for which we find that it typically traces gas at $A_V\approx{}6.1_{-1.0}^{+1.2}~\rm{}mag$, corresponding to a characteristic $\rm{}H_2$ density $\approx{}870~\rm{}cm^{-3}$ (Sec.~\ref{sec:cd-characteristic}). The only molecular transition clearly connected to dense gas is the $\rm{}N_2H^+$ (1--0) transition, characteristic of  $A_V\approx{}16~\rm{}mag$ and densities $\approx{}4,000~\rm{}cm^{-3}$. The low characteristic densities derived for the HCN (1--0) line are about two orders of magnitude below values commonly adopted in extragalactic research (Sec.~\ref{sec:hcn-density}). This impacts theoretical discussions of SF trends in galaxies. We use this new knowledge on the emission from HCN to compare SF in galaxies to SF in the Milky Way (Sec.~\ref{sec:sf-relations}). The comparisons indicate that galaxies either deviate from SF relations holding in the Milky Way, or hitherto unknown reservoirs of emission contribute to $L_{\text{HCN (1--0)}}$.

\begin{acknowledgements}
We thank a knowledgeable anonymous referee for helping to significantly improve the paper. This research was conducted in part at the Jet Propulsion Laboratory, which is operated by the California Institute of Technology under contract with the National Aeronautics and Space Administration (NASA). AG acknowledges support from Fondecyt under grant 3150570.
\end{acknowledgements}

\bibliographystyle{aa}
\bibliography{/Users/jens/texinputs/mendeley/library}

\end{document}